\documentclass[reprint,superscriptaddress,nofootinbib, amsmath,amssymb, aps,prl]{revtex4-1}

\usepackage[utf8]{inputenc}
\usepackage{graphicx}
\usepackage{bm}
\usepackage{amsmath,amsthm}
\usepackage{amsfonts,amssymb}
\usepackage{amsfonts}
\usepackage{enumitem}
\usepackage[dvipsnames]{xcolor}
\usepackage{hyperref}
\usepackage{tcolorbox}
\usepackage{chngcntr}
\usepackage[toc, page,titletoc]{appendix}
\usepackage{latexsym,epsfig,bbm}

\def\bra#1{\langle{#1}|}
\def\ket#1{|{#1}\rangle}

\newcommand{\figref}[1]{Fig.~\ref{#1}}

\newtheorem{thm}{Theorem}
\newtheorem{prof}{Proof}
\newtheorem{lem}{Lemma}
\newtheorem{col}{Corollary}

\begin{document}

\title{Measures of distinguishability between stochastic processes}

\begin{abstract}
	Quantifying how distinguishable two stochastic processes are lies at the heart of many fields, such as machine learning and quantitative finance. While several measures have been proposed for this task, none have universal applicability and ease of use.  In this Letter, we suggest a set of requirements for a well-behaved measure of process distinguishability. Moreover, we propose a family of measures, called  divergence rates, that satisfy all of these requirements. 
	Focussing on a particular member of this family -- the co-emission divergence rate -- we show that it can be computed efficiently, behaves qualitatively similar to other commonly-used measures in their regimes of applicability, and remains well-behaved in scenarios where other measures break down.	
\end{abstract}

\author{Chengran Yang}
\email{Yangchengran92@gmail.com}
\affiliation{
	School of Physical and Mathematical Sciences, Nanyang Technological University, Singapore 637371
}
\affiliation{Complexity Institute, Nanyang Technological University, Singapore 637335}

\author{Felix C.~Binder}
\email{quantum@felix-binder.net}
\affiliation{
	Institute for Quantum Optics and Quantum Information - IQOQI Vienna, Austrian Academy of Sciences, Boltzmanngasse 3, 1090 Vienna, Austria
}

\author{Mile Gu}
\email{mgu@quantumcomplexity.org}
\affiliation{
	School of Physical and Mathematical Sciences, Nanyang Technological University, Singapore 637371
}
\affiliation{Complexity Institute, Nanyang Technological University, Singapore 637335}
\affiliation{
	Centre for Quantum Technologies, National University of Singapore, 3 Science Drive 2, Singapore 117543
}

\author{Thomas J.~Elliott}
\email{physics@tjelliott.net}
\affiliation{Complexity Institute, Nanyang Technological University, Singapore 637335}
\affiliation{
	School of Physical and Mathematical Sciences, Nanyang Technological University, Singapore 637371
}

\date{\today}

\maketitle


How alike are the behaviours of two systems? How similar are the trajectories of two stock prices? Much of the physical world can be described as a collection of interacting stochastic processes; understanding how distinguishable two processes are allows us to answer such questions. These problems are of universal relevance -- for example, quantifying how closely a model replicates its target has applications in fields such as protein homology~\cite{soding2004protein} and speech recognition~\cite{rabiner1989tutorial,silva2008upper}. Meanwhile,  understanding how much external noise or perturbations impact upon the behaviour of a system is a central task in studies of open systems~\cite{horsthemke1984noise}, quantum computation~\cite{preskill2018quantum}, and machine learning~\cite{kennewick2012system,Huang2013}.

Many measures have been proposed for this task~\cite{hellinger1909neue, kullback1951information, bures1969extension, juang1985probabilistic, geometry1986adventure, lyngso1999metrics, Jenson-Shannon, cha2007comprehensive, sahraeian2011novel}. However, a number of these are founded on measures tailored for quantifying distances between distributions. Though they work well for quantifying distances between finite strings, they typically do not behave well in the context of processes where infinite strings of observational data arise as a process continues to run. Particularly, they fail to quantify \emph{how} different processes are~\cite{hellinger1909neue, bures1969extension, Jenson-Shannon, geometry1986adventure, lyngso1999metrics}. Others, such as the Kullback-Leibler (KL)-divergence~\cite{kullback1951information,juang1985probabilistic}, require intensive computational resources to evaluate and can behave pathologically in seemingly innocuous situations such as processes with different output alphabets. Finally, measures that possess an intrinsic dependence on a representation of a process rather than the process itself will often fail by misidentifying different models of the same process with identical observable behaviour as being distinguishable~\cite{levinson1983introduction, sahraeian2011novel}.

In this Letter, we suggest several requirements a measure of process distinguishability should satisfy. We then propose a family of measures that satisfy all of these properties. We focus on a specific member of this family and develop an efficient method to compute the exact value of this measure. Furthermore, we illustrate our proposal by applying it to a set of example scenarios designed to highlight where other measures either cannot be applied or behave pathologically.


Consider a bi-infinite, discrete-time, discrete-alphabet stochastic process $\mathcal{P}$, which generates an output $x$ drawn from an alphabet $\mathcal{A}$ at each time step. A contiguous output sequence $x_{t:t+L}:= x_t x_{t+1}\cdots x_{t+L-1}$ occurs with probability $P(x_{t:t+L})$, where $t$ denotes the initial time and $L$ is the length of sequence. Many naturally occurring processes can be described within this formalism, such as biological processes~\cite{Baldi1059,Barrett1998} and speech recognition~\cite{Juang1991,Schuller2003}. We shall here consider processes that are both stationary and ergodic: a process is stationary if the distribution of its output sequences are invariant with respect to time, i.e., $P(x_{t:t+L})=P(x_{0:L})\forall t,L\in\mathbb{Z},x_{0:L}\in\mathcal{A}^L$; a process is ergodic if its time-average behaviour is identical to its ensemble-average -- and its statistical properties can be deduced from a single sufficiently long sample of an output sequence.


With the above questions as motivation, we suggest that a good measure of distinguishability between stochastic processes $R(\mathcal{P},\mathcal{Q})$ should satisfy the following criteria:
\begin{enumerate}

\item \emph{Non-negativity}: $R(\mathcal{P},\mathcal{Q})\geq 0$.

\item \emph{Symmetry}: $R(\mathcal{P},\mathcal{Q})=R(\mathcal{Q},\mathcal{P})$.

\item \emph{Identity of indiscernibles}: $R(\mathcal{P},\mathcal{Q})=0\Leftrightarrow \mathcal{P} = \mathcal{Q}$, i.e., $R(\mathcal{P},\mathcal{Q})=0$ iff the processes are identical.

Together these three conditions define a semimetric distance. Note that as with many proposed measures of distance between processes, we do not demand the triangle inequality $R(\mathcal{P},\mathcal{Q})\leq R(\mathcal{P},\mathcal{G}) + R(\mathcal{Q},\mathcal{G})\forall\mathcal{G}$ be satisfied, and so the measure will not necessarily be a metric.

\item \emph{Model-independence}: $R(\mathcal{P},\mathcal{Q})$ should depend only on observable properties (i.e., the outputs) of the processess, and not any underlying models.

This condition enforces that the measure should be identical when calculated relative to any representation of the processes, alleviating the issues discussed above for model-dependent measures.

\item \emph{Continuity}: Suppose stochastic process $\mathcal{Q}$ depends on a continuous parameter $\delta$. Continuity mandates that $\lim_{\delta \to \delta_0} R(\mathcal{P},\mathcal{Q}(\delta))= R(\mathcal{P},\mathcal{Q}(\delta_0))$. 

Smooth deformations in the parameters defining a process will smoothly change the distinguishability between it and other processes -- this condition enforces that this is reflected in the measure. 
\end{enumerate}


With these requirements at hand, we are able to assess the behavior of existing measures that have been proposed for quantifying how distinct two stochastic processes are. One such widely-used measure, often serving as a cost function for many machine learning works, is the aforementioned KL-divergence
\begin{equation}
D_{KL}(P||Q) = \sum_{x_{t:t+L}} P(x_{t:t+L})\log\left[ \frac{P(x_{t:t+L})}{Q(x_{t:t+L})}\right],
\end{equation}
where here and throughout logarithms are of base $2$. It can be seen however, that the KL-divergence does not satisfy the continuity criterion. It becomes singular when two stochastic processes contain different sets of possible output sequences, no matter how small the probability is of these unique sequences occuring. Moreover, while not a violation of any of the above criteria, a further drawback of the KL-divergence is its high computational cost, as it requires a calculation over all output sequences, the number of which grows exponentially with their length $L$.

Other measures such as the Jensen-Shannon-divergence~\cite{Jenson-Shannon}, though free from singularities, can still fail the continuity criterion in the context of processes where the output sequence lengths are infinite. This problem is endemic to measures based on distances between distributions, such as trace norms and the Bures distance~\cite{bures1969extension} -- and highlights a crucial key difference between distributions over finite sequences and processes. Specifically, as any two different processes can be asymptotically distinguished for sufficiently long output sequences, for such sequences these measures asymptotically saturate to their maximal value -- thus identifying whether $\mathcal{P}$ and $\mathcal{Q}$ are different processes, but not how different they are. A good measure of process distinguishability should be equipped to handle this distinction.  

Finally, the model independence criterion rules out other measures~\cite{levinson1983introduction,sahraeian2011novel} that are explicitly based on the structure of a particular model. That is, for such measures two models of the same stochastic process may be identified as having non-zero distance between them despite exhibiting identical observable behaviour. Furthermore, such model-dependent measures may also be impossible to evaluate for some pairs of models with sufficiently distinct structures. 


In light of such issues with commonly-used measures, we seek a measure that satisfies all of our criteria. We propose a family of measures of process distinguishability, called divergence rates, which measure how quickly the observed behaviour of two processes becomes distinguishable. That is, they quantify how much the distance between output sequence probability distributions grows with their length. 

Consider a metric distance measure between distributions $D(P,Q)$ that is normalised such that $0\leq D(P,Q)\leq1$. We introduce the notion of \emph{similarity} $S_D$, that can be thought of as the complement to the distance, satisfying
\begin{equation}
S_D(P,Q):=\sqrt{1-D(P,Q)^2}
\end{equation}
We then define the $D$ \emph{divergence rate} as
\begin{equation}
\label{eqddr}
R_D(\mathcal{P},\mathcal{Q}):=-\lim_{L\to\infty}\frac{1}{L}\log[S_D(P,Q;L)],
\end{equation}
where $S_D(P,Q;L)$ and similarly $D(P,Q;L)$ are used to denote these quantities evaluated for distributions formed from sequences of length $L$ output by the processes. The $D$ divergence rate can be seen to parameterise the rate at which the similarity (according to the distance $D$) of the two processes decays once many symbols have been observed. This parallels the notion of entropy rates~\cite{cover2012elements}, which similarly capture long-term behaviour by averaging over long sequences.

\begin{thm}
$R_D(\mathcal{P},\mathcal{Q})$ fulfils the above requirements on a measure of process distinguishability for any continuous, normalised metric distance $D(P,Q;L)$ that scales with $L$ as $D(P,Q;L)\sim 1-\alpha\exp(-\eta L)$ for non-negative real $\eta$ and non-negative $\alpha\leq1$. For such a distance, we have that $R_D(\mathcal{P},\mathcal{Q})=\eta/2$.
\end{thm}	

\begin{prof}
	Conditions 1, 2 and 4 immediately follow from the definition of $R_D(\mathcal{P},\mathcal{Q})$ and the properties of $D(P,Q)$ as a metric. By directly inserting the scaling $D(P,Q;L)=1-\alpha\exp(-\eta L)$ into Eq.~\eqref{eqddr} it can be seen that $R_D(\mathcal{P},\mathcal{Q})=\eta/2$. As $D(P,Q)$ is a metric it follows that if two processes are identical we must have $\eta=0$, and conversely $\eta=0$ indicates that two processes have identical long-term behaviour -- and hence condition 3 is satisfied. Note that $\alpha$ is irrelevant, and depends only on transient behaviour resulting from the initial configuration of the two processes. Finally, condition 5 also follows from directly inserting the scaling into Eq.~\eqref{eqddr}. 
\end{prof}

Furthermore, whenever $D(P,Q)$ exhibits the required scaling, we see that $R_D(\mathcal{P},\mathcal{Q})$ is infinite iff $D(P,Q;L)$ becomes exactly 1 within finite $L$, rather than just asymptotically approaching it. Qualitatively, this can be understood as the measure being infinite iff the two processes can be discriminated with certainty by observing a sufficiently long yet finite sequence of outputs.


We now consider the case where the distance used is the $L_2$ norm, given by
\begin{equation}
	D_{L_2}(P,Q):=\frac{1}{\sqrt{2}}\lVert \hat{P}-\hat{Q}\rVert_2
\end{equation}
where $\hat{P}=P/\lVert P\rVert_2$ and $\hat{Q}=Q/\lVert Q\rVert_2$. Noting that $\lVert \hat{P} -\hat{Q}\rVert_2^2 = 2- 2\langle\hat{P},\hat{Q}\rangle$, where $\langle\cdot,\cdot\rangle$ is the inner product, this distance can be expressed in terms of so-called co-emission probabilities~\cite{lyngso1999metrics} $C(P,Q;L)=\sum_{x_{0:L}}P(x_{0:L})Q(x_{0:L})$. Using this, we obtain
\begin{equation}
R_C(\mathcal{P},\mathcal{Q}) = -\lim_{L\to \infty}\frac{1}{2L}\log\left[\frac{C(P,Q;L)}{\sqrt{C(P,P;L)C(Q,Q;L)}}\right],
\end{equation}
which we call the \emph{co-emission divergence rate} (CDR).

\begin{thm}
	The CDR satisfies all of the requirements specified for a good measure of process distinguishability.
\end{thm}

As $D_{L_2}$ is a continuous metric distance, we need only tp show that the measure obeys the specified scaling.

\begin{lem}
	$D_{L_2}(P,Q;L)$ scales as $1-\alpha\exp(-\eta L)$
\end{lem}

The proof of this employs a recently-developed correspondence between tensor networks and stochastic processes~\cite{Yang2018}, and is given in detail in Supplementary Information B.

Any bi-infinite, stationary stochastic process can be represented in terms of a hidden Markov model (HMM)~\cite{Shalizi2001}. Such models consist of a set of hidden internal states $s_i$. At each time step, based on the current state $s_i$ the model generates output $x$ with probability $P(x|s_i)$ and transitions to state $s_j$ with probability $P(s_j|s_i,x)$.  In proving Lemma 1, we obtain an efficient way to compute the CDR between any two processes for which we know a HMM representation.

\begin{col}
	Given a HMM representation of process $\mathcal{P}$ with transition probabilities $P(s_j,x|s_i)$ and of process $\mathcal{Q}$ with $Q(\tilde{s}_n,x|\tilde{s}_m)$, the CDR between them is given by
	\begin{equation}
	R_C(\mathcal{P},\mathcal{Q})= -\frac{1}{2}\log\left[ \frac{\mu_{PQ}}{\sqrt{\mu_{P}}\sqrt{\mu_{Q}}}\right],\label{Eq: co-emiision divergence rate evaluation}
	\end{equation}
	where $\mu_{P}$, $\mu_{Q}$ and $\mu_{PQ}$ are the leading eigenvalues of the transfer matrices $\mathbb{E}_{PP}$, $\mathbb{E}_{QQ}$ and $\mathbb{E}_{PQ}$, defined as 
	\begin{equation}
		(\mathbb{E}_{PQ})_{im,jn} :=\sum_x P(s_j,x|s_i)Q(\tilde{s}_n,x|\tilde{s}_m)
	\end{equation}
\end{col}

The computational complexity of calculating these eigenvalues (and hence the CDR) depends only on the number of hidden states $|S|$ in our HMM representations of $\mathcal{P}$ and $\mathcal{Q}$, scaling polynomially with both. We need only calculate the  leading eigenvalue of a $|S_\mathcal{P}||S_\mathcal{Q}|\times|S_\mathcal{P}||S_\mathcal{Q}|$ matrix for $\mu_{PQ}$, and similarly for $\mu_P$ and $\mu_Q$. Moreover, as we only need the leading eigenvalues, we can use tools such as the power method~\cite{mises1929praktische} rather than full spectral decomposition. Crucially, there is no scaling of complexity with the length of sequences considered (the $L\to\infty$ limit is implicitly accounted for). And unlike Monte Carlo methods used to estimate e.g. KL-divergences, the result is exact.

\begin{figure}[htp]
	\includegraphics[scale=0.5]{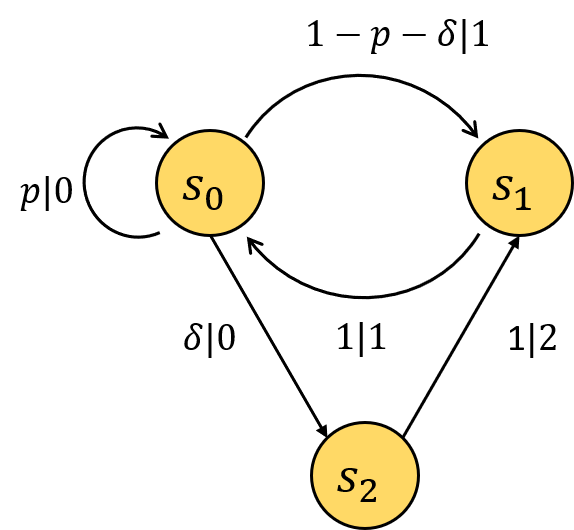}
	\caption{The example process we consider can be represented by a three state HMM with two variables $p$ and $\delta$. The edge label $P|x$ between states $s_i$ and $s_j$ signifies that if the model is in state $s_i$ it will transition to $s_j$ whilst emitting symbol $x$ with probability $P$.}

\label{Fig:Three states}
\end{figure}

In Supplementary Information C we work through a pedagogical example that demonstrates how our efficient method for calculating the CDR as described in Corollary 1 may be used. Here in the main text we present an illustrative example using a highly tunable process that highlights several scenarios in which our measure can be employed, where other previously proposed measures of process distinguishability break down. The most general form of this example process can be represented by a HMM with three hidden states, as illustrated in \figref{Fig:Three states}. The model has two variable parameters $p$ and $\delta$; we use $\mathcal{G}(p,\delta)$ to represent the process generated by the model for a particular set of parameters

First, we show that the CDR exhibits qualitatively similar behaviour to the (symmetric) KL-divergence (per symbol) in a scenario where the latter can be applied. Note that we must utilise Monte-Carlo methods~\cite{hershey2007approximating} to estimate the KL-divergence, due to its computationally-intesive nature.
Let  process $\mathcal{P}=\mathcal{G}(p,0)$ for $p\in[0.1,0.9]$, and similarly process $\mathcal{Q}=\mathcal{G}(q,0)$ for $q\in[0.1,0.9]$. We calculate the CDR using the method described in Corollary 1, while the symmetric KL-divergence per symbol is estimated using the Monte Carlo method for sequences of length $L=1000$ and a sampling set size $M=50$.
\begin{figure}[htp]
	\includegraphics[scale=0.3]{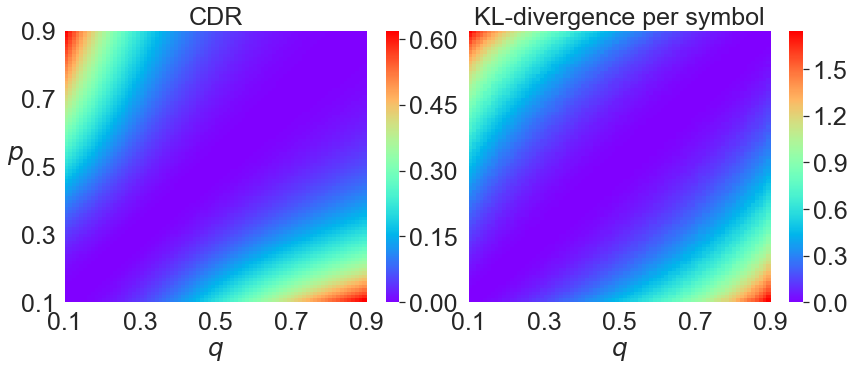}
	\caption{Comparison of CDR and KL-divergence per symbol for distinguishing between different parameter values of the process at $\delta=0$. We see that the qualitative behaviour of the two is very similar. As would be expected, both are zero along the line $p=q$ where the processes are equal, and grow as the difference $|p-q|$ increases.}
	\label{Fig: TS_2}
\end{figure}

From \figref{Fig: TS_2}, we see that the CDR and KL-divergence exhibit similar behavior.  We emphasise that we are able to efficiently compute the exact CDR, while we are only able to estimate the KL-divergence as it requires exponentially growing resources with sequence length. Moreover, as the processes considered have infinite Markov order (i.e. their behaviour is conditioned on outputs from infinitely far back into the past), no measure based on sequences with finite $L$ can capture the full behaviour of the processes exactly.

\begin{figure}[htp]
	\centering
	\includegraphics[scale=0.80]{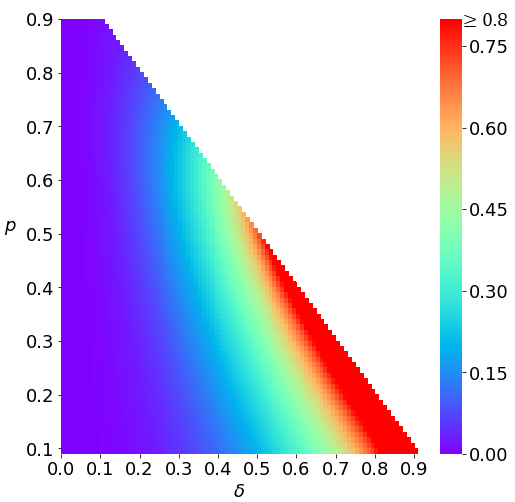}
	\caption{The CDR is able to be calculated to ascertain the distinguishability between processes with different output alphabets and representations with different numbers of states, unlike other measures such as the KL-divergence. As expected, we see that the CDR increases as one process becomes increasingly likely to emit a symbol the other cannot. The white region represents an unphysical parameter regime.}
	\label{Fig: TS_1}
\end{figure}

Second, we consider a scenario where the KL-divergence cannot be suitably used. Consider the case where again process $\mathcal{P}=\mathcal{G}(p,0)$ for $p\in[0.1,0.9]$, but now $\mathcal{Q}=\mathcal{G}(q_1,q_2)$ for $q_1\in[0.1,0.9]$ and $q_2\in[0,1-q_1]$. When $\delta=0$ the hidden state $s_2$ cannot be reached, and so the symbol 2 is never emitted -- thus for any $q_2\neq0$, $\mathcal{P}$ and $\mathcal{Q}$ have different output alphabets and so exhibit infinite KL-divergence (per symbol). Nevertheless, the CDR varies smoothly with the parameters, and we are still able to efficiently calculate it, as shown in Fig.~\ref{Fig: TS_1} for the plane defined by $p=q_1$. Furthermore, since the HMM representations of the processes have different numbers of accessible states for $q_2\neq0$, a number of other measures based on the model topology cannot be properly applied~\cite{Levinson1983}.

Though our efficient method for computing the CDR relies on having HMM representations of the processes considered, the measure itself does not rely on this. In lieu of HMM representations, Monte Carlo methods can be employed on the sequence probabilities to calculate the CDR, as with the KL-divergence. We also note that while we have here considered stationary processes, a modified form of the CDR can be applied to non-stationary processes, where instead of taking $L\to\infty$ we take the longest sequence possible; the measure will then yield the average decay of process similarity per symbol.

A further generalisation that can be considered is the effect of relabelling the alphabet,  such that each output is ascribed to a different symbol. In general, the observed statistics after such a relabelling will be different, and as such the CDR between the original and relabelled processes will generally be non-zero. However, from another perspective, one can argue that the two processes still exhibit statistically identical behaviour -- with merely a different nomenclature for the events. To remedy this, one can consider an alphabet-symmetrised form of the divergence rate, where it is minimised over all permutations of the alphabet for one of the processes. This would then identify a process and its relabelled version as having zero CDR.

Finally, we remark that there exist other members of the divergence rate family which satisfy all of the requirements for a process distinguishability measure. Consider the Bures~\cite{bures1969extension} or  Hellinger distance~\cite{hellinger1909neue} $D_B(P,Q;L) = \sqrt{1 - F(P,Q;L)}$, where the fidelity $F(P,Q;L) := \sum_{x_{0:L}}\sqrt{P(x_{0:L})Q(x_{0:L})}$. Taking this as our distance measure, we obtain the \emph{fidelity divergence rate} (FDR):
\begin{equation}
	R_F(\mathcal{P},\mathcal{Q}) = -\lim_{L\to \infty} \frac{1}{2L} \log_2 [F(P,Q;L)].
\end{equation}
In Supplementary Information D we show that the FDR satisfies all the requirements, and provide an efficient way to calculate it from deterministic HMM representations of the processes.


To summarise, we have discussed a set of conditions we believe a good measure of processs distinguishability should satisfy, and proposed a family of divergence rates that satisfy them. We focussed on a particular example of this family, the CDR, and developed an efficient method for computing it. We illustrate the advantages of our measure relative to previously proposed measures by applying it to example scenarios where other measures behave pathologically. Finally, we discussed a number of possibile generalisations of the measure.

Our measure can be applied to a broad range of areas, particularly those dealing with stochastic processes such as HMMs~\cite{Levinson1983}, computational mechanics~\cite{Crutchfield1989,Shalizi2001} and quantum stochastic modelling~\cite{gu2012quantum, mahoney2016occam, aghamohammadi2018extreme, elliott2018superior, binder2018practical,elliott2019memory, liu2019optimal}. Other areas of application include assessment of the accuracy of machine learning models, and quantifying the robustness of processes to noise. Our method for efficiently computing the CDR uses tools from tensor networks~\cite{perez2006matrix, Orus2007,Temme2010,June2014}, adding to the growing list of applications of these methods for stochastic processes~\cite{hieida1998application, carlon1999density, carlon2001critical, critch2014algebraic, Kliesch2014, johnson2015capturing, Yang2018} and machine learning~\cite{oseledets2011tensor, Stoudenmire2016, han2018unsupervised, stoudenmire2018learning, Guo2018, Chen2018, Clark_2018, glasser2018supervised, Levine2019}.

\acknowledgements
This work was funded by the Lee Kuan Yew Endowment Fund (Postdoctoral Fellowship), grant FQXi-RFP-1809 from the Foundational Questions Institute and Fetzer Franklin Fund (a donor advised fund of Silicon Valley Community Foundation), Singapore Ministry of Education Tier 1 grant RG190/17, and National Research Foundation Fellowship NRF-NRFF2016-02. F.C.B. acknowledges funding from the European Union’s Horizon 2020 research and innovation programme under the Marie Skłodowska‐Curie grant agreement No 801110 and the Austrian Federal Ministry of Education, Science and Research (BMBWF). T.J.E., C.Y.,~and F.C.B.~thank the Centre for Quantum Technologies for their hospitality. 

\bibliography{Reference}


\pagebreak
\clearpage
\widetext
\appendix

\begin{center}
\textbf{\large Supplementary Information - A measure of distinguishability between stochastic processes}
\end{center}

\begin{center}
Chengran Yang$,^{1,\;2,\;*}$ Felix C.~Binder$,^{3,\;1,\;2,\;\dagger}$ Mile Gu$,^{1,\;2,\;4,\;\ddag}$ and Thomas J.~Elliott${}^{2,\;1,\;\mathsection}$

\vspace{0.2em}
\emph{\small ${}^{\mathit{1}}$School of Physical and Mathematical Sciences, Nanyang Technological University, Singapore 637371\\
${}^{\mathit{2}}$Complexity Institute, Nanyang Technological University, Singapore 637723\\
${}^{\mathit{3}}$Institute for Quantum Optics and Quantum Information - IQOQI Vienna,\\ Austrian Academy of Sciences, Boltzmanngasse 3, 1090 Vienna, Austria\\
${}^{\mathit{4}}$Centre for Quantum Technologies, National University of Singapore, 3 Science Drive 2, Singapore 117543}\\
{\small (Dated: \today)}
\end{center}

\setcounter{equation}{0}
\setcounter{figure}{0}
\setcounter{table}{0}
\setcounter{page}{1}
\renewcommand{\theequation}{S\arabic{equation}}
\renewcommand{\thefigure}{S\arabic{figure}}
\renewcommand{\thepage}{S\arabic{page}}

In this Supplementary Information, we (A) introduce the relevant background of tensor networks, and show their connection to hidden Markov models (HMMs). Using these tools, we then (B) provide the proofs of Theorem 2, Lemma 1, and Corollary 1 of the main paper. We (C) provide a pedagogical example of using our efficient method to calculate the co-emission divergence rate (CDR). Finally, we (D) show that the fidelity divergence rate (FDR) also satisfies our requirements.


\section{A: Tensor networks and their relation to HMMs}

\begin{figure}[htp]
\includegraphics[scale=0.4]{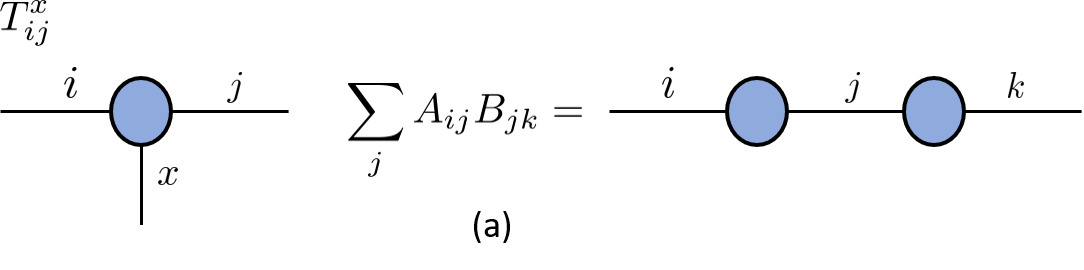}\\
\includegraphics[scale=0.4]{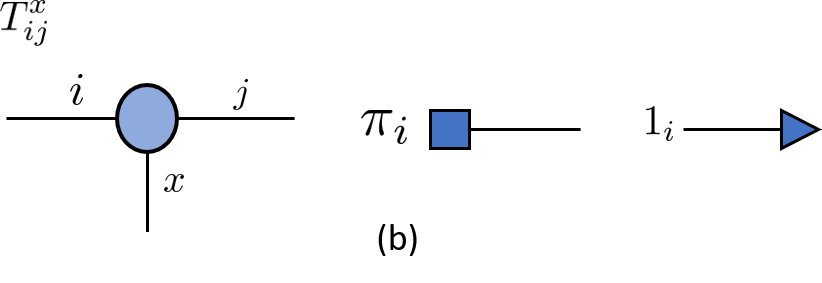}\quad\quad\quad\quad\quad
\includegraphics[scale=0.4]{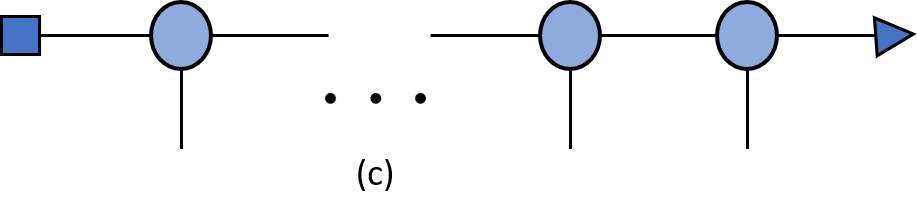}
	\caption{Pictorial representation of tensor networks. (a) Each leg represents an index of a tensor, with linking between legs representing summation over the corresponding index. (b) Tensor network representation of a HMM. (c) Multi-index tensor representing the probability of an output sequence.}
\label{App:Fig: Tensor representation}
\end{figure}

A tensor network decomposes a large tensor into several smaller tensors connected by a network structure. These techniques have many promising applications, a key one being in simplifying the numerical simulation of quantum many-body systems. They possess a comprehensive pictorial representation in which each tensor is represented by a node with several legs, as shown in \figref{App:Fig: Tensor representation}(a). 

To represent a HMM we can use a special type of tensor network, called matrix product states~\cite{Yang2018}. Transitions between states in the HMM are described by the transition matrix $T^x_{ij} := P(s_j,x|s_i)$; this is a rank $3$ tensor, and is thus represented by a node with three legs as shown in \figref{App:Fig: Tensor representation}(b). The stationary distribution of the HMM states $\pi_i$ is represented by the square, and the triangle represents $1_i$, a column vector filled with $1$s.

The probability of a particular sequence $x_{0:L}$ being generated by a HMM is given by 
\begin{equation}
P(x_{0:L}) = \sum_{s_i^0,s_i^1\cdots s_i^{L-1}} \pi_i^0P(s_i^1,x_0|s_i^0)\cdots P(s_i^{L},x_{L-1}|s_i^{L-1}).
\end{equation}
This can be represented by a tensor network, as shown in \figref{App:Fig: Tensor representation}(c). The nomenclature `matrix product state' becomes clear: the sequence tensor is obtained by multiplying by a matrix $T^{x_l}$ at each step. 

A HMM, and its tensor network representation, decompose the large tensor $P(x_{0:L})$ into products of small tensors $T^x_{ij}$. As a result, HMMs exponentially reduce the memory requirement of representing a stochastic process to $O(LN^2)$ from $O(|\mathcal{A}|^L)$ where $L$ is the length of the sequence, $N$ is the number of states of the HMM and $|\mathcal{A}|$ is the size of output alphabet. 


\section{B: Proof of Theorem 2, Lemma 1, and Corollary 1}
\label{App: Theorem: co-emission divergence rate.}
\begin{figure}[htp]
\includegraphics[scale=0.4]{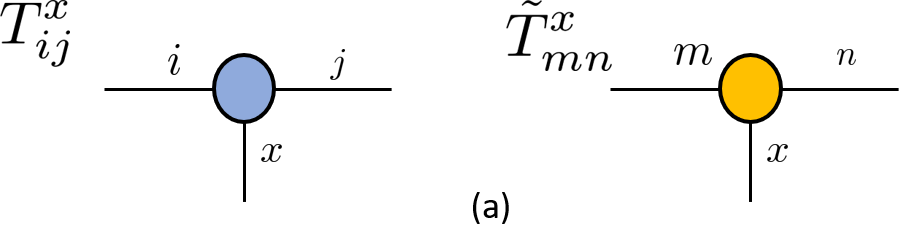}\quad\quad\quad\quad\quad\quad
\includegraphics[scale=0.4]{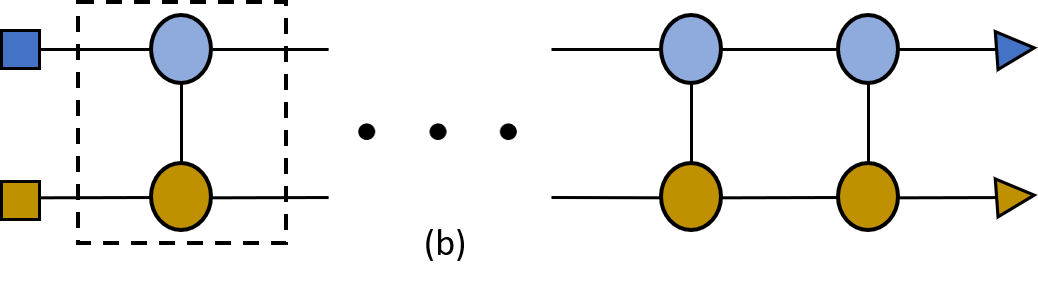}
	\caption{(a) Tensor network representation of the transition matrices of $\mathcal{P}$ and $\mathcal{Q}$. (b) Tensor network representation of the co-emission probability.}
\label{App: Fig:Transition tensors}
\end{figure}

Here, we present our efficient method of computing the co-emission divergence rate (CDR), in the process proving Theorem 2 and Lemma 1. Every stationary stochastic process has a HMM representation~\cite{Shalizi2001}; we consider two stationary stochastic processes $\mathcal{P}$ and $\mathcal{Q}$ with HMMs $T^x_{ij}:=P(s_j,x|s_i)$ and $\tilde{T}^x_{mn}:=Q(\tilde{s}_n,x|\tilde{s}_m)$, where $s_i$ and $\tilde{s}_m$ are the corresponding hidden states. The corresponding pictorial representations are shown in \figref{App: Fig:Transition tensors}(a).

The co-emission probability is 
\begin{equation}
	C(P,Q;L) = \sum_{x_{t:t+L}} P(x_{t:t+L})Q(x_{t:t+L}),
\end{equation}
obtained by contracting the output indices $x_{t:t+L}$ over tensors $T$ and $\tilde{T}$, as shown in the Fig.~\ref{App: Fig:Transition tensors}(b). The tensor structure in the dashed square, which repeatedly appears in the network, has four legs, i.e., is a rank $4$ tensor. Combining the left two legs together as a row index, and the right two legs as a column index, this becomes a matrix -- the transfer matrix:
\begin{equation}
	(\mathbb{E}_{PQ})_{im,jn} :=\sum_x P(s_j,x|s_i)Q(\tilde{s}_n,x|\tilde{s}_m).
\end{equation} 
The leftmost and rightmost tensors represent the left and right boundaries, respectively. The left boundary $\langle bl|$ is a row vector with elements $v_{ij}=\pi_i\tilde{\pi}_j$. The right boundary $|br\rangle$ is a column vector filled with $1$s, such that the hidden states at the last step are equally weighted, i.e., $P(x_{t:t+L})Q(x_{t:t+L})=\sum_{s_j,s_n}P(x_{t:t+L},s_j)Q(x_{t:t+L},s_n)$.

If $\mathbb{E}_{PQ}$ is diagonalisable,  it has an eigenvalue decomposition
\begin{equation}
	\mathbb{E}_{PQ} = \sum_i \mu_i \ket{r_i}\bra{l_i}
\end{equation} 
where $\mu_i$ are eigenvalues of $\mathbb{E}_{PQ}$, sorted in order of decreasing magnitude, and $\ket{r_i}$ and $\bra{l_i}$ are the associated right- and left-eigenvectors. Consequently, we have 
\begin{equation}
	\mathbb{E}_{PQ}^L = \mu_1^L (\ket{r_1}\bra{l_1} +  \sum_{i\neq1} (\frac{\mu_i}{\mu_1})^L \ket{r_i}\bra{l_i}) 
\end{equation}
As $\mathbb{E}_{PQ}$ is constructed from probabilities, it is non-negative. Its left- and right-leading eigenvectors are then non-negative according to the Perron-Frobenius theorem~\cite{perron1907theorie, frobenius1912matrizen}. Thus, the left- and right-boundary vectors have non-zero overlap with the associated leading left- and right-eigenvectors of the matrix $\mathbb{E}_{PQ}$, and therefore the co-emission probability has the following scaling
\begin{equation}
C(P,Q;L)  = \langle bl|\mathbb{E}_{PQ}^L |br\rangle = \mu_{PQ}^L(\alpha_{PQ} + O((\frac{\mu_2}{\mu_{PQ}})^L)) \sim \alpha_{PQ}\mu_{PQ}^L
\end{equation}
where $\mu_{PQ} :=\mu_1$ and $\alpha_{PQ} = \bra{bl}r_1\rangle\bra{l_1}br\rangle$ is positive. This scaling holds even if $\mathbb{E}_{PQ}$ is not diagonalisable; this can be proved using the Jordan form of the matrix. 

Using the same argument, we also have
\begin{equation}
		C(P,P;L) \sim \alpha_P\mu_{P}^L\quad\quad\quad\mathrm{and}\quad\quad\quad C(Q,Q;L) \sim \alpha_Q\mu_{Q}^L.
\end{equation}
where $\mu_P$ is the leading eigenvalue of the transfer matrix $\mathbb{E}_{PP}$ and $\mu_{Q}$ is the leading eigenvalue of the transfer matrix $\mathbb{E}_{QQ}$. Then, we have 
\begin{equation}
		D_{L_2}(P,Q;L)^2 = 1 - \frac{C(P,Q;L)}{\sqrt{C(P,P;L)C(Q,Q;L)}}\sim 1- \alpha (\frac{\mu_{PQ}}{\sqrt{\mu_P\mu_Q}})^L
\end{equation}
where $\alpha = \alpha_{PQ}/\sqrt{\alpha_P\alpha_Q}$. Thus, the distance has the desired scaling, proving Lemma 1, and in turn Theorem 2. Taking $L\to\infty$ leads to
\begin{equation}
\lim_{L\to \infty} \frac{1}{L}\log (\sum_{x_{t:t+L}} P(x_{t:t+L})Q(x_{t:t+L})) = \log \mu_{PQ},
\end{equation}
and analogously, 
\begin{align}
\lim_{L\to \infty} \frac{1}{L}\log (\sum_{x_{t:t+L}} P(x_{t:t+L})P(x_{t:t+L})) &= \log \mu_{P}\\
\lim_{L\to \infty} \frac{1}{L}\log (\sum_{x_{t:t+L}} Q(x_{t:t+L})Q(x_{t:t+L})) &= \log \mu_{Q}.
\end{align}
Therefore, 
\begin{equation}
	R_C(\mathcal{P},\mathcal{Q}) = -\frac{1}{2}\log\frac{\mu_{PQ}}{\sqrt{\mu_{P}\mu_{Q}}}.
\end{equation}
This proves Corollary 1.


\section{C: Pedagogical example of calculating the CDR}
\label{App: Analytical example: Perturbed coin}

\begin{figure}[htp]
	\includegraphics[scale=0.5]{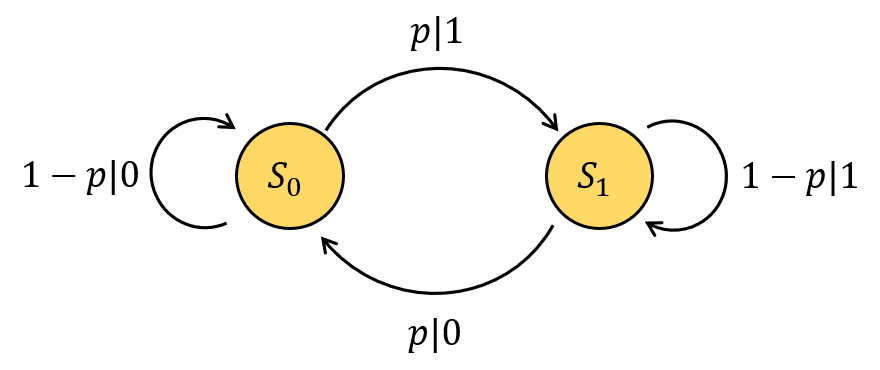}
	\caption{The perturbed coin process has two hidden states, $s_0$ and $s_1$. The system occupies $s_0$ when the last output was $0$, and similarly, $s_1$ after output $1$. }\label{App:Fig:Perturbed Coin}
\end{figure} 

As a pedagogical example of how our efficient method for computing the CDR works, we study the distinguishability between two versions of the perturbed coin process~\cite{gu2012quantum} (representable by the HMM in \figref{App:Fig:Perturbed Coin}) with different parameters. This is a Markov process, as output $0$ indicates the hidden state is $s_0$ and output $1$ indicates the hidden state $s_1$.  Consider two perturbed coin processes $\mathcal{P}$ and $\mathcal{Q}$ with parameters $p$ and $q$ respectively. Then the transfer matrices are 
\begin{equation}
	\mathbb{E}_{PP} = \begin{bmatrix}
		(1-p)^2 & 0 & 0 & p^2\\
		(1-p)p & 0 & 0 & (1-p)p\\
		(1-p)p & 0 & 0 & (1-p)p\\
		p^2 & 0 & 0 & (1-p)^2		
	\end{bmatrix}\quad
	\mathbb{E}_{QQ} = \begin{bmatrix}
		(1-q)^2 & 0 & 0 & q^2\\
		(1-q)q & 0 & 0 & (1-q)q\\
		(1-q)q & 0 & 0 & (1-q)q\\
		q^2 & 0 & 0 & (1-q)^2		
	\end{bmatrix}\quad
	\mathbb{E}_{PQ} = \begin{bmatrix}
		(1-p)(1-q) & 0 & 0 & pq\\
		(1-p)q & 0 & 0 & p(1-q)\\
		p(1-q) & 0 & 0 & (1-p)q\\
		pq & 0 & 0 & (1-p)(1-q)		
	\end{bmatrix}\quad
\end{equation}
Evaluating the leading eigenvalues of these matrices, we obtain
\begin{equation}
	\mu_{P} = p^2+(1-p)^2, \quad \mu_Q = q^2+(1-q)^2,\quad \mu_{PQ} = pq +(1-p)(1-q)
\end{equation}

Therefore, the CDR is
\begin{equation}
R_C(\mathcal{P},\mathcal{Q}) = -\frac{1}{2}\log_2 \frac{pq +(1-p)(1-q)}{\sqrt{(p^2+(1-p)^2)(q^2+(1-q)^2)}}
\end{equation}
Clearly, $R(\mathcal{P},\mathcal{Q}) = 0$ iff the two processes are identical, i.e., $p=q$.


\section{D: Fidelity divergence rate}
\label{App: Theorem: fidelity divergence rate}

Here, we present another member of the divergence rate family that also satisfies the desired properties of a process distinguishability measure. This divergence rate is called fidelity divergence rate (FDR), as the associated distance is the Bures/Hellinger distance $D_B(P,Q;L) = \sqrt{1-F(P,Q;L)}$, which is expressed in terms of the fidelity $F(P,Q;L):=\sum_{x_{t:t+L}}\sqrt{P(x_{t:t+L})Q(x_{t:t+L})}$.
Then $S_{D_B}(P,Q;L) = \sqrt{1-D_B(P,Q;L)^2} = \sqrt{F(P,Q;L)}$,  and the FDR is 
\begin{equation}
	R_F(\mathcal{P},\mathcal{Q}) = -\frac{1}{2} \lim_{L\to\infty} \frac{1}{L}\log(F(P,Q;L))
\end{equation}

Similar to the CDR, we will demonstrate that $D_B$ exhibits the required scaling for the FDR to satisfy our requirements, and provide an efficient method for its evaluation given deterministic HMM  representations of the processes. A deterministic (or unifilar) HMM is one for which the current hidden state can always be deduced with certainty given the previous state and output. This means that each state only has at most one outgoing edge for each symbol. Every stationary stochastic process has a deterministic HMM representation~\cite{Shalizi2001}.

\begin{thm}
	The FDR satisfies all the proposed requirements for a measure of process distinguishability. 
\end{thm}

Consider two stationary stochastic processes, $\mathcal{P}$ and $\mathcal{Q}$, with deterministic HMM representations $ P(s_j,x|s_i)$ and $Q(\tilde{s}_n,x|\tilde{s}_m)$, respectively, and associated transfer matrix $(\mathbb{E}^F_{PQ})_{im,jn} = \sum_x \sqrt{P(s_j|x,s_i)Q(\tilde{s}_n,x|\tilde{s}_m)}$.

\begin{lem}
	$D_{B}(P,Q;L)$ scales as $1-\alpha\exp(-\eta L)$
\end{lem}

\begin{figure}[htp]
\includegraphics[scale=0.4]{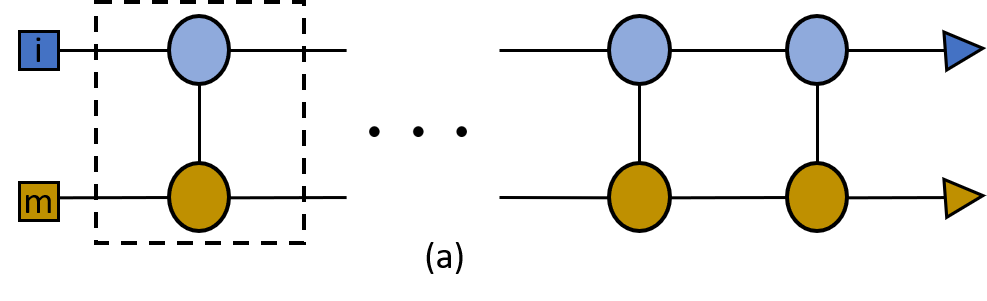}\quad\quad\quad\quad\includegraphics[scale=0.4]{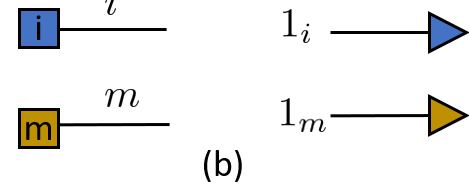}
\caption{Tensor network representation of (a) the fidelity and (b) boundary vectors.}
\label{App:Fig: F_boundary}
\end{figure}

The fidelity $\sum_{x_{0:L}}\sqrt{P(x_{0:L}|s_i)Q(x_{0:L}|\tilde{s}_m)}$, conditioned on starting in hidden states $(s_i,\tilde{s}_m)$ has a pictorial representation as shown in \figref{App:Fig: F_boundary}(a). The left boundary represents the $i^{\mathrm{th}}$ and $m^{\mathrm{th}}$ standard basis vectors $\langle i|$ and $\langle m|$ in the corresponding space, while the right boundary, denoted by $|br\rangle$, is the column vector filled with $1$s, as shown in Fig~\ref{App:Fig: F_boundary}(b).

The tensor structure in the dashed square is the transfer matrix $\mathbb{E}^F_{PQ}$, which acts repeatedly on the left boundary $\langle i,m|$. Similar to the proof for the co-emission, if $\mathbb{E}_{PQ}^F$ is diagonalisable we have the eigenvalue decomposition
	\begin{equation}
		\mathbb{E}_{PQ}^F = \sum_i \mu_i \ket{r_i}\bra{l_i}
	\end{equation} 
	where $\mu_i$ are the eigenvalues of $\mathbb{E}_{PQ}^F$ sorted in order of decreasing magnitude, and $\ket{r_i}$ and $\bra{l_i}$ are the associated right- and left-eigenvectors. Consequently,
	\begin{equation}
		(\mathbb{E}_{PQ}^F)^L = \mu_1^L (\ket{r_1}\bra{l_1} +  \sum_{i\neq1} (\frac{\mu_i}{\mu_1})^L \ket{r_i}\bra{l_i}) 
	\end{equation}
    where $\alpha = \bra{bl}r_1\rangle\bra{l_1}br\rangle$ is the overlap between the left vector and the leading eigenvector of $\mathbb{E}_{PQ}^F$. Because $\mathbb{E}_{PQ}$ is non-negative matrix, its left- and right-leading eigenvectors are non-negative according to the Perron-Frobenius theorem~\cite{perron1907theorie,frobenius1912matrizen}. Since $\langle i,m|$ spans the whole space, there always exists a vector  $\langle i,m|$ such that it has non-zero overlap with leading left eigenvector of transfer operator $\mathbb{E}^F_{PQ}$, i.e., $\langle i,m|r_1\rangle > 0$. As with the co-emission, the above scaling still holds when $\mathbb{E}^F_{PQ}$ is not diagonalisable, as can be shown using the Jordan form. Thus, we see that the fidelity decays exponentially with the length of the sequence, and thus $D_B$ exhibits the required scaling, proving Lemma 2 and Theorem 3.

\begin{col}
	Given deterministic HMM representations $ P(s_j,x|s_i)$ and $Q(\tilde{s}_n,x|\tilde{s}_m)$ of processes $\mathcal{P}$ and $\mathcal{Q}$, the FDR is given by
	\begin{equation}
	R_F(\mathcal{P},\mathcal{Q}) = -\frac{1}{2}\log_2 \mu_{PQ}
	\end{equation}
	where $\mu_{PQ}$ is the leading eigenvalue of operator $\mathbb{E}^F_{PQ} = \sum_x \sqrt{P(s_j|x,s_i)Q(\tilde{s}_n,x|\tilde{s}_m)}$. 
\end{col}

For all boundary vectors $\langle i,m|$, we have 
	
	\begin{equation}
		\lim_{L\rightarrow\infty}{\frac{1}{L}\log{\sum_{x_{0:L}}\sqrt{P\left(x_{0:L}\middle| s_i\right)Q\left(x_{0:L}\middle| \tilde{s}_m\right)}}}=\log{\mu_{PQ}}
	\end{equation}
	where $\mu_{PQ} = \mu_1$
	The above quantity is the fidelity conditional on certain past $(s_i,\tilde{s}_m)$. We now bound the non-conditioned fidelity in the following.
	\begin{equation}
	\begin{split}
	F(P,Q;L) := &\sum_{x_{0:L}} \sqrt{P(x_{0:L})Q(x_{0:L})}\\ &=\sum_{x_{0:L}}\sqrt{\sum_{i}\pi_{i}P\left(x_{0:L}\middle| s_i\right)}\times \sqrt{\sum_{m}\tilde{\pi}_{m}Q\left(x_{0:L}\middle| \tilde{s}_m\right)}
	\end{split}
	\end{equation}
	The inequality $\sqrt{x+y}\leq \sqrt{x}+\sqrt{y}$ implies
	\begin{equation}
	\begin{split}
	F(P,Q;L)&\leq \sum_{x_{0:L},i,m}\sqrt{\pi_{i}\tilde{\pi}_m} \times\sqrt{P\left(x_{0:L}\middle| s_i\right)Q\left(x_{0:L}\middle| \tilde{s}_m\right)}\\
	&\leq \max_{i,m} (\sum_{x_{0:L}} \sqrt{P\left(x_{0:L}\middle| s_i\right)Q\left(x_{0:L}\middle| \tilde{s}_m\right)})
	\end{split}
	\end{equation}
	Therefore, we have 
	\begin{equation}
	\begin{split}
		R_F(\mathcal{P},\mathcal{Q}) &= \frac{1}{2}\lim_{L\to \infty} -\frac 1L \log_2 F(P,Q;L) \\
		&\geq \frac{1}{2}\lim_{L\to \infty} -\frac 1L \log_2 \max_{i,m} (\sum_{x_{0:L}} \sqrt{P\left(x_{0:L}\middle| s_i\right)Q\left(x_{0:L}\middle| \tilde{s}_m\right)}) \\
		&= -\frac{1}{2}\log_2 \mu_{PQ}
	\end{split}
	\end{equation}
	On the other hand, using inequality $\sqrt{\sum_i p_i x_i}\geq\sum_i p_i\sqrt{x_i}$, we have
	\begin{equation}
	\begin{split}
	F(P,Q;L)\geq & \sum_{i,m} \pi_{i}\tilde{\pi}_{m}\sum_{x_{0:L}} \sqrt{P\left(x_{0:L}| s_i\right)Q\left(x_{0:L}| \tilde{s}_m\right)}\\
	&\geq \max_{i,m} \pi_{i}\tilde{\pi}_{m}\sum_{x_{0:L}} \sqrt{P\left(x_{0:L}| s_i\right)Q\left(x_{0:L}| \tilde{s}_m\right)}
	\end{split}
	\end{equation}
	Similarly, we also obtain
	\[ R_F(P,Q) \leq -\frac{1}{2}\log_2\mu_{PQ}\]
	Thus, the proof is completed.

The fidelity divergence rate can thus be obtained by evaluating the leading eigenvalue of the transfer matrix $\mathbb{E}_{PQ}^F$. The computational complexity of this method depends only polynomially on the number of hidden states in the deterministic HMM representations of each process, and thus can be efficiently computed.

Finally, we provide upper and lower bounds for the FDR that can be calculated even when we do not have deterministic representations of the processes, from their statistics alone.
\begin{thm}
	Suppose two stochastic processes $\mathcal{P},\mathcal{Q}$ have finite Markov order and the larger one is $\kappa$. Then the fidelity divergence rate has the following upper and lower bounds
	\begin{equation}
		\begin{split}
		R^{\downarrow}&:= \min_{x_{-\kappa:0}}-\log F[P(x|x_{-\kappa:0}),Q(x|x_{-\kappa:0})] \leq 2R_F(\mathcal{P},\mathcal{Q})\\
		R^{\uparrow}&:= \max_{x_{-\kappa:0}}-\log F[P(x|x_{-\kappa:0}),Q(x|x_{-\kappa:0})] \geq 2R_F(\mathcal{P},\mathcal{Q})
		\end{split}
	\end{equation}
\end{thm}

	First, having Markov order $\kappa$ implies that 
	\begin{equation}
	\begin{split}
	P(x_{0:L+\kappa+1}) &= P(x_{L+\kappa}|x_{0:L+\kappa}) P(x_{0:L+\kappa})\\
	& = P(x_{L+\kappa}|x_{L:L+\kappa})P(x_{0:L+\kappa})
	\end{split}
	\end{equation}
	From this, we find that 
	\begin{equation}
	P(x_{0:L+\kappa+1}) \leq P(x_{0:L+\kappa}) \max_{x_{L:L+\kappa}}P(x_{L+\kappa}|x_{L:L+\kappa})
	\end{equation}
	Thus, we have
	\begin{equation}
	\begin{split}
	&F(P,Q;L+\kappa+1)\\
	&= \sum_{x_{0:L+\kappa+1}} \sqrt{P(x_{0:L+\kappa+1})Q(x_{0:L+\kappa+1})}\\
	&= \sum_{x_{0:L+\kappa}}\sqrt{P(x_{0:L+\kappa})Q(x_{0:L+\kappa})} \times\sum_{x_{L+\kappa}}\sqrt{P(x_{L+\kappa}|x_{L:L+\kappa})Q(x_{L+\kappa}|x_{L:L+\kappa})}\\
	&\leq \sum_{x_{0:L+\kappa}}\sqrt{P(x_{0:L+\kappa})Q(x_{0:L+\kappa})} \times\max_{x_{L:L+\kappa}}\sum_{x_{L+\kappa}}\sqrt{P(x_{L+\kappa}|x_{L:L+\kappa})Q(x_{L+\kappa}|x_{L:L+\kappa})}\\
	&= F(P,Q;L+\kappa) \times \max_{x_{L:L+\kappa}}\sum_{x_{L+\kappa}}\sqrt{P(x_{L+\kappa}|x_{L:L+\kappa})Q(x_{L+\kappa}|x_{L:L+\kappa})}.
	\end{split}
	\end{equation}
	Substituting the above into the definition of FDR leads to
	\begin{equation}
	\begin{split}
	2R_F(\mathcal{P},\mathcal{Q}) &= \lim_{L\to \infty} -\frac 1L\log_2(F(P,Q;L))\\
	&\leq \lim_{L\to \infty} -\frac 1L(\log_2F(P,Q;\kappa)+ (L-\kappa)R^{\uparrow})\\
	&=R^{\uparrow}
	\end{split}
	\end{equation}
	The lower bound $2R_F(\mathcal{P},\mathcal{Q})\leq R^\downarrow$  can similarly be obtained by replacing maximisations with minimisations and reversing the directions of the inequalities.

\end{document}